\documentclass[twocolumn,showpacs,preprintnumbers,amsmath,aps,prl]{revtex4}

\usepackage{graphicx}% Include figure files
\usepackage{graphics}% Include figure files
\usepackage{dcolumn}% Align table columns on decimal point
\usepackage{bm}% bold math

\begin{document}
\preprint{}

\title{Magnetic Structure and Interactions in the Quasi-1D Antiferromagnet CaV$_2$O$_4$}

\author{O.~Pieper$^{1,2}$}
\email[]{oliver.pieper@helmholtz-berlin.de}
\author{B.~Lake$^{1,2}$}
\author{A.~Daoud-Aladine$^{3}$}
\author{M.~Reehuis$^{1,4}$}
\author{K.~Proke$\check{\text{s}}^{1}$}
\author{B.~Klemke$^{1}$}
\author{K.~Kiefer$^{1}$}
\author{J.Q.~Yan$^{5}$}
\author{A.~Niazi$^{5}$}
\author{D.C.~Johnston$^{5}$}
\author{A.~Honecker$^{6}$}

%\affiliation{$^{1}$Helmholtz-Zentrum Berlin f\"{u}r Materialien und Energie (HZB), Glienicker Stra\ss e 100, 14109 Berlin, Germany}
%\affiliation{$^{2}$Institut f\"{u}r Festk\"{o}rperphysik, Technische Universit\"{a}t Berlin, Hardenbergstra\ss e 36, 10623 Berlin, Germany}
%\affiliation{$^{3}$ISIS Facility, Rutherford Appleton Laboratory, Chilton, Didcot, Oxon OX11 0QX, UK}
%\affiliation{$^{4}$Max-Planck-Institut f\"{u}r Festk\"{o}rperforschung, Heisenbergstr. 1, D-70569 Stuttgart, Germany}
%\affiliation{$^{5}$Ames Laboratory and Department of Physics and Astronomy, Iowa State University, Ames, Iowa 50011, USA}
%\affiliation{$^{6}$Universit\"{a}t G\"{o}ttingen, Institut f\"{u}r Theoretische Physik, D-37077 G\"{o}ttingen, Germany}

\affiliation{$^{1}$Helmholtz-Zentrum Berlin f\"{u}r Materialien und Energie (HZB), Glienicker Stra\ss e 100, 14109 Berlin, Germany\\
$^{2}$Institut f\"{u}r Festk\"{o}rperphysik, Technische Universit\"{a}t Berlin, Hardenbergstra\ss e 36, 10623 Berlin, Germany\\
$^{3}$ISIS Facility, Rutherford Appleton Laboratory, Chilton, Didcot, Oxon OX11 0QX, United Kingdom\\
$^{4}$Max-Planck-Institut f\"{u}r Festk\"{o}rperforschung, Heisenbergstr. 1, D-70569 Stuttgart, Germany\\
$^{5}$Ames Laboratory and Department of Physics and Astronomy, Iowa State University, Ames, Iowa 50011, USA\\
$^{6}$Institut f\"{u}r Theoretische Physik, Universit\"{a}t G\"{o}ttingen, D-37077 G\"{o}ttingen, Germany}

\begin{abstract}
CaV$_2$O$_4$ is a spin-1 antiferromagnet, where the magnetic vanadium ions have an orbital degree of freedom and are arranged on quasi-one-dimensional zig-zag chains. The first- and second-neighbor vanadium separations are approximately equal suggesting frustrated antiferromagnetic exchange interactions. High-temperature susceptibility and single-crystal neutron diffraction measurements are used to deduce the dominant exchange paths and orbital configurations. The results suggest that at high temperatures CaV$_2$O$_4$ behaves as a Haldane chain, but at low temperatures, it is a spin-1 ladder. These two magnetic structures are explained by different orbital configurations and show how orbital ordering can drive a system from one exotic spin Hamiltonian to another.
\end{abstract}


\maketitle

\date{\today}
%\cite{QMbook,Ramirez94,Greedan00}
%Introduction
The magnetism of low-dimensional and frustrated systems is a current research topic. Often such phenomena arise from the crystal structure, e.g., well-separated chains or planes of magnetic ions can lead to low-dimensional magnetism while triangular structures combined with antiferromagnetic interactions give rise to frustration (e.g., Ref. \cite{QMbook}). More recently the role of orbitals on exchange paths has been investigated and materials where direct exchange is the dominant magnetic interaction are particularly sensitive to orbital occupation and overlap. In some cases the structure gives rise to orbital configurations and exchange interactions that could not be deduced from a simple inspection of magnetic ion separations. Recent studies of vanadium spinels suggest, that geometrically frustrated systems with an orbital degree of freedom can relax frustration via orbital ordering which gives rise to preferred exchange pathways \cite{Lee04,Tsunetsugu03,Tchernyshyov04,Zhang06,Pardo}. In some cases there is a lowering of the dimensionality of the magnetism with dominant antiferromagnetic interactions coupling the magnetic moments into spin-1 chains (Haldane chains) \cite{Lee04,Tsunetsugu03} or into pairs (dimers) \cite{Pardo}. Here we investigate CaV$_2$O$_4$ where the vanadium ions interact predominantly via direct exchange and are arranged on zig-zag chains. The intrachain V-V distances are approximately equal, suggesting competing first- and second-neighbor antiferromagnetic exchange interactions. However, we find that the magnetism is more complex than this simple picture suggests and is strongly influenced by the orbital ordering of the vanadium ions.\\\indent
\begin{figure}[htb!]
        \includegraphics[width=7.0cm]{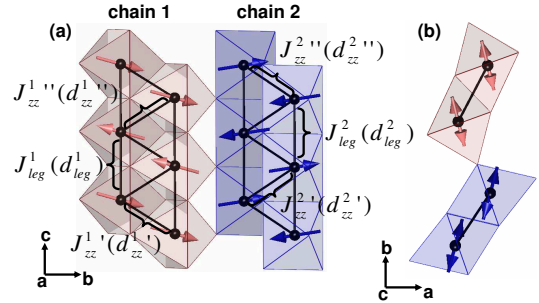}
        \caption{(Color online). The double chain structure of CaV$_2$O$_4$ projected onto the (a) \textit{b-c} and (b) \textit{a-b} planes. The two inequivalent VO$_6$ octahedra are highlighted in pink (chain 1) and blue (chain 2). Arrows indicate the direction of the ordered moments and the intrachain exchange constants for the monoclinic phase are labeled. In the orthorhombic phase ${J^{n}_{\text{zz}}}'={J^{n}_{\text{zz}}}''={J^{n}_{\text{zz}}}$, $J^{1}_{\text{zz}}\approx J^{2}_{\text{zz}}$, $J^{1}_{\text{leg}}\approx J^{2}_{\text{leg}}$.}
    \label{fig:Pic1_zigzag_chains}
\end{figure}
Unlike many similar AV$_2$O$_4$ systems that are spinels, CaV$_2$O$_4$ crystallizes in the CaFe$_2$O$_4$-type structure with orthorhombic space group $Pnam$ at room temperature. The magnetic V$^{3+}$ ions possess spin $S=1$ due to two unpaired electrons in the $3d$ shell and are arranged in two inequivalent (although similar) zig-zag chains of slightly distorted edge-sharing VO$_6$ octahedra running along the crystallographic $c$ direction (see Fig. \ref{fig:Pic1_zigzag_chains}). 

The octahedral environment partially removes the degeneracy of the $3d$ orbitals and both electrons occupy the lower lying $t_{2g}$ levels. The combination of edge-sharing octahedra which allows the overlap of $t_{2g}$ orbitals and V-V distances of $\approx3$ \AA\ favors direct exchange interactions \cite{Rogers62}. Such interactions are antiferromagnetic and highly sensitive to both orbital occupation and vanadium separation \cite{Rogers62,Blanco07}. Superexchange via oxygen can also contribute to the coupling; however, the V-O-V intrachain bond angles range from 92.5$^{\circ}$ to 99.3$^{\circ}$, implying that these interactions are weak. The structure suggests chains consisting of two competing antiferromagnetic interactions, $J^{n}_{\text{zz}}$ along the zig-zags and $J^{n}_{\text{leg}}$ along the legs where the superscript $n$ takes the values 1 or 2 for the two inequivalent chains. The V-V distances along the zig-zags and legs are the same for the two chains with $d^{1}_{\text{zz}}=d^{2}_{\text{zz}}=3.08$ \AA\ and $d^{1}_{\text{leg}}=d^{2}_{\text{leg}}=3.01$ \AA\, and the exchange constants should also be similar $J^{1}_{\text{zz}}\approx J^{2}_{\text{zz}}$ and $J^{1}_{\text{leg}}\approx J^{2}_{\text{leg}}$. The interchain coupling is weak and frustrated and occurs via superexchange interactions with V-O-V angles ranging from 122.0$^{\circ}$ to 131.2$^{\circ}$. This material is therefore potentially a quasi-one-dimensional (1D), spin-1 antiferromagnet with frustrated first ($J_{\text{zz}}$) and second ($J_{\text{leg}}$) neighbor interactions.

CaV$_2$O$_4$ undergoes a structural phase transition at $T_s\approx141$ K to a monoclinic phase with space group $P2_1/n$ ($P2_1/n11$) resulting in $\alpha = 90.767(1)^{\circ}$ and a small change in lattice parameters \cite{Yan08}. Below  $T_s$, the zig-zag distances become inequivalent ($d^{n}_{\text{zz}}{'}\ne d^{n}_{\text{zz}}{''}$), which lifts the degeneracy of the interactions along the zig-zags so that at low temperatures, $J^{n}_{\text{zz}}{'}\ne J^{n}_{\text{zz}}{''}$ (see Fig. \ref{fig:Pic1_zigzag_chains}). However, a single distance $d^{n}_{\text{leg}}$ and interaction ($J^{n}_{\text{leg}}$) remain along the legs. The monoclinic distortion changes the V-V and V-O distances in slightly different ways for the two chains, therefore the set of interactions characterizing each chain are probably similar but not identical (see supplementary information \cite{supp1} and \cite{Yan08}). 

Below $T_N\approx 51-78$ K, CaV$_2$O$_4$ develops long-range antiferromagnetic order with a magnetic propagation vector $\bm{k}=(0,\frac{1}{2},\frac{1}{2})$ \cite{Hastings67,Bertaut67,Zong07Prb,Niazi08,Sugiyama08}. Experimental investigations of powder and single crystal samples using various techniques found that the ordered spin moment had a strongly reduced value of $1.0\mu_B\leq\mu\leq1.59\mu_B$ (compared to 2 $\mu_B$ available for $S=1$) and that the average spin direction is along the $b$-axis \cite{Hastings67,Bertaut67,Zong07Prb,Niazi08,Sugiyama08}. The reduced ordered moment as well as a broad maximum in the dc susceptibility at $T\approx 250-300$ K \cite{Zong07Prb,Niazi08,Kikuchi01} indicate low dimensional and/or frustrated behavior. Neutron powder diffraction experiments find a collinear spin structure at low temperatures \cite{Hastings67,Bertaut67}. In contrast, nuclear magnetic resonance (NMR) measurements indicate that the structure consists of two collinear substructures that are canted with respect to each other \cite{Zong07Prb}. 

\begin{figure}[t!]
        \includegraphics[width=7.0cm]{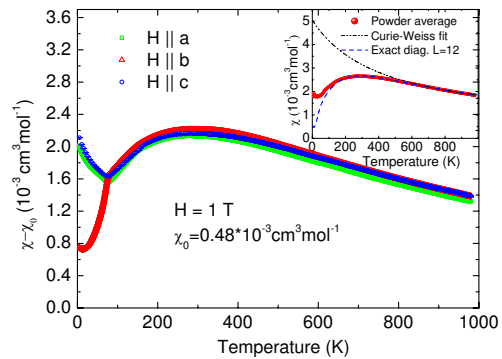}
        \caption{(Color online). Single-crystal dc magnetic susceptibility of CaV$_2$O$_4$ for all three crystallographic directions. The inset shows the powder averaged data fitted to Curie-Weiss and exact diagonalization models.}
    \label{fig:Pic5_suscept}
\end{figure}

Herein we discuss the first single-crystal neutron diffraction and high temperature dc susceptibility measurements on CaV$_2$O$_4$ using the samples described in Ref. \cite{Niazi08}. Remarkably the susceptibility indicates that the material behaves as a $S=1$ Heisenberg chain above $T_s$ instead of a zig-zag chain with $J_{\text{leg}}\approx J_{\text{zz}}$ as anticipated from the crystal structure. 
Neutron diffraction is used to determine the magnetic structure at 6 K and reveals that below $T_s$ the exchange interactions are those of a $S=1$ ladder. 
These results can be explained by two different orbital configurations at high and low temperatures and imply that orbital ordering drives the system between the two spin Hamiltonians.\\ 
The static magnetic susceptibility of CaV$_2$O$_4$ was measured over a temperature range 3 $\leq T\leq$ 1000 K using a vibrating sample magnetometer (Quantum Design) at the Laboratory for Magnetic Measurements, HZB. Data were collected on samples of volume $\approx6$ mm$^3$ with a magnetic field of 1 T applied along the crystallographic axes. The background corrected susceptibility is shown in Fig. \ref{fig:Pic5_suscept}. The low-temperature data are in agreement with previous measurements \cite{Zong07Prb, Niazi08}. The high-temperature data (700 $\leq T\leq$ 950 K) were fitted to the Curie-Weiss law giving a $g$ factor of 1.958(4) and a paramagnetic Curie temperature of $\Theta=-418(5)$ K. This yields an effective moment of $\mu_{\text{exp}}=2.77(18)\mu_B$, as commonly found in $V^{3+}$ materials. In addition the sum of the magnetic exchange couplings in the orthorhombic phase, $J_{\text{zz}}+J_{\text{leg}}=-(3/2)k_B\Theta[S(S+1)]^{-1}=27.0(4)$ meV, was obtained assuming identical chains.

To extract further information about the high-temperature exchange interactions in CaV$_2$O$_4$, exact diagonalization calculations for a spin-1 chain (up to 12 spins) with nearest-neighbor ($J_{\text{zz}}$) and next-nearest-neighbor ($J_{\text{leg}}$) interactions were performed \cite{ALPS}. This model was fitted to the data in the orthorhombic phase (200 $\leq T\leq$ 950 K) and the best agreement was found for $J_{\text{zz}}=19.85(1)$ meV and $J_{\text{leg}}=0.75(4)$ meV (inset of Fig. \ref{fig:Pic5_suscept}), while a second solution was found for $J_{\text{zz}}=3.02(1)$ meV and $J_{\text{leg}}=18.60(8)$ meV. These results show that one of the exchange interactions is much stronger than the other, rather than both being approximately equal and frustrated as expected from the crystal structure. The two solutions are in fact magnetically equivalent since either $J_{\text{zz}}$ is dominant (single spin-1 chain) or $J_{\text{leg}}$ is dominant (two spin-1 chains). 

%Magnetic neutron diffraction
Single-crystal neutron diffraction experiments were performed to obtain the low-temperature magnetic structure of CaV$_2$O$_4$ using the instruments SXD at the ISIS facility, Rutherford Appleton Laboratory, and E4 and E5 at the BER II reactor, HZB, Germany. 
Only the E5 measurements are reported here. E5 is a four-circle single-crystal diffractometer with a two-dimensional position-sensitive $^{3}$He-detector. A pyrolytic graphite monochromator selected an incident wavelength of $\lambda=2.36$ \AA\ and the cylindrical sample $(d=4$ mm, $h=6$ mm) was cooled by a closed-cycle refrigerator. Data were collected above the structural and magnetic phase transitions at 160 K and in the antiferromagnetically ordered phase at 6 K.

\begin{figure}[htb!]
		\includegraphics[width=6.0cm]{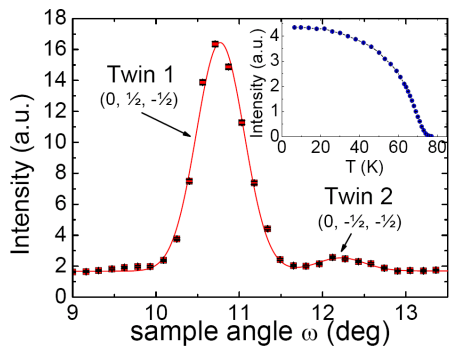}
		\caption{(Color online). Magnetic intensity as a function of sample angle $\omega$, at the orthorhombic $(0,\frac{1}{2},\frac{\bar{1}}{2})$ position for $T=6$ K. The temperature dependence of the integrated magnetic Bragg intensity of twin 1 $(0,\frac{1}{2},\frac{\bar{1}}{2})$ is shown in the inset.}
	\label{fig:Twins_Fcalc}
\end{figure}

The orthorhombic-monoclinic structural transition gives rise to two twin domains occupying equal volume fractions. This is observed as split contributions at the nominal Bragg positions of the orthorhombic cell with $l\neq0$. Figure \ref{fig:Twins_Fcalc} shows the magnetic Bragg intensity at 6 K for $(0,\frac{1}{2},\frac{\bar{1}}{2})$ (orthorhombic notation). The two peaks could be indexed in the monoclinic setting as $(0,\frac{1}{2},\frac{\bar{1}}{2})$ from twin 1 and $(0,\frac{\bar{1}}{2},\frac{\bar{1}}{2})$ from twin 2, and the large difference in their intensities is due to the difference in their magnetic structure factors. It is important for the magnetic refinement to accurately determine the intensity of the twin peaks separately; however, since the monoclinic splitting is relatively small, most of the twin pairs partially overlap, making it impossible to use conventional integration routines. Therefore manual fitting and correction was done, and by applying the twin law $(h,k,l)_{\text{Twin 1}}\sim (h,-k,l)_{\text{Twin 2}}$, it was possible to identify the peaks and create a separate intensity list for each. In addition, a list containing the summed intensities of both twins was created for the peaks which could not be separated. Note that due to the resolution function of neutron powder diffractometers the splitting cannot be observed in the powder measurements \cite{Hastings67,Bertaut67}; thus the twinned peaks are summed resulting in a significant loss of information.

The magnetic structure refinement was carried out with these three lists simultaneously using the program \texttt{FULLPROF} \cite{Fullprof}.
As a starting point we used the two collinear solutions proposed from the previous neutron powder-diffraction work which were compatible with symmetry analysis \cite{Hastings67, Bertaut67}. They provide good fits to our single crystal data with magnetic residual factors of $R_F=0.13$ and $0.15$ respectively. A close inspection, however, reveals that they systematically fail to assign a non-zero intensity to the small contributions such as the $(0,\frac{\bar{1}}{2},\frac{\bar{1}}{2})_{\text{Twin 2}}$ peak (Fig. \ref{fig:Twins_Fcalc}). We therefore considered canted versions of these models and found much better agreement. The best solution has a magnetic residual factor of $R_F=0.048$ and is illustrated in Fig. \ref{fig:Pic1_zigzag_chains}. 

The spins are found to be collinear within each zig-zag chain with antiparallel alignment along the legs and alternating antiparallel-parallel alignment along the zig-zags. The two inequivalent chains are canted with respect to each other by equal and opposite amounts from the $b$ axis. The total canting angle between the chains is 37(2)$^{\circ}$ with projections in the $a$-$b$ and $b$-$c$ planes of 29(1)$^{\circ}$ and 24(2)$^{\circ}$, respectively. The refined magnetic moment sizes per V atom are strongly reduced from the expected value $\langle \mu \rangle=gS\mu_B\approx2$ $\mu_B$, with 0.97(1) $\mu_B$ and 1.01(1) $\mu_B$ for the two chains, respectively, consistent with the value of 1.06(6) $\mu_B$ obtained from the powder diffraction \cite{Hastings67, Bertaut67}. NMR measurements revealed the existence of two antiferromagnetic substructures, canted about the $b$ axis \cite{Zong07Prb} by a total canting angle of 19(1)$^{\circ}$ and the ordered spin moment was found to lie in the range $1.02-1.59$ $	\mu_B$.
Unlike single-crystal neutron diffraction, however, NMR is unable to determine which spins form the two substructures or the relative ordering within them. Here we show that each substructure is a zig-zag chain.

To gain further insight into the magnetism of CaV$_2$O$_4$ we examine the orbital occupation and direct exchange mechanism more deeply. The three $t_{2g}$ orbitals have lobes pointing along the three intrachain directions $d^{n}_{\text{leg}}$, $d^{n}_{\text{zz}}{'}$ and $d^{n}_{\text{zz}}{''}$ respectively which correspond to the three exchange paths $J^{n}_{\text{leg}}$, $J^{n}_{\text{zz}}{'}$ and $J^{n}_{\text{zz}}{''}$. As there are only two electrons in the $3d$ shell, it is important to know which orbitals are occupied since only filled orbitals give rise to strong antiferromagnetic exchange interactions. The degeneracy of the $t_{2g}$ orbitals can be lifted by distortions in the VO$_6$ octahedra and an examination of the V-O distances can be used to deduce the orbital energy level diagram. 
In CaV$_2$O$_4$ those structural distortions are less pronounced than for classical Jahn-Teller systems with $e_g$ orbital order [e.g., LaMnO$_3$ (Ref. \cite{Caravajal97})]. However, magnetism and orbital physics in $t_{2g}$ systems are very sensitive to structural changes and even small distortions can have a dramatic impact on the physics \cite{Blake02, Reehuis06PRB}.

\begin{figure}[htb!]
        \includegraphics[width=7.0 cm]{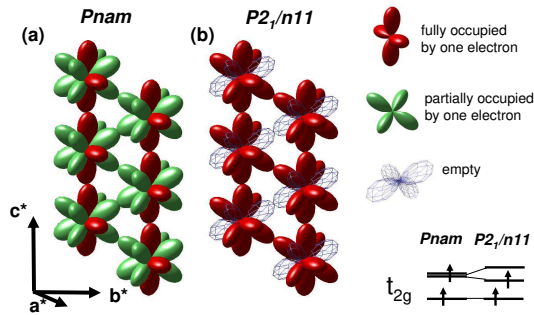}
        \caption{(Color online). The orientation and occupancy of the vanadium $t_{2g}$ orbitals for an example chain in the (a) high temperature ($Pnam$) phase and (b) low temperature ($P2_1/n11$) phase. (Lower right) $t_{2g}$ energy level diagram for both phases.}
    \label{fig:Pic4_mag_structure_bc}
\end{figure}

In the high-temperature orthorhombic phase of CaV$_2$O$_4$ the octahedra are compressed and the $d^{n}_{\text{leg}}$ orbital (with lobes along $d^{n}_{\text{leg}}$) is shifted to lower energy while the higher-energy $d^{n}_{\text{zz}}{'}$ and $d^{n}_{\text{zz}}{''}$ orbitals are degenerate and symmetrically equivalent \cite{supp1}. The $d^{n}_{\text{leg}}$ orbital is therefore occupied by one of the two electrons leading to a strong antiferromagnetic leg interaction $J^{n}_{\text{leg}}$, while the remaining electron will partially occupy the two degenerate $d^{n}_{\text{zz}}$ orbitals so that the zig-zag interactions are identical ($J^{n}_{\text{zz}}{'}=J^{n}_{\text{zz}}{''}$) and significantly weaker than $J^{n}_{\text{leg}}$ \cite{Onoda03}. The predicted high temperature orbital states are illustrated in Fig.\ \ref{fig:Pic4_mag_structure_bc}(a). Our susceptibility analysis reveals that one of the intrachain exchanges is much stronger than the other although it is unable to distinguish which; this orbital arrangement combined with the slightly shorter $d^{n}_{\text{leg}}=3.01$ \AA\ compared to $d^{n}_{\text{zz}}=3.08$ \AA\ suggests that the dominant interaction is $J^{n}_{\text{leg}}$.

The orthorhombic-to-monoclinic transition at $T_{s}=141$ K gives rise to two inequivalent zig-zag distances ($d^{n}_{\text{zz}}{'}$ and $d^{n}_{\text{zz}}{''}$) and exchange interactions ($J^{n}_{\text{zz}}{'}$ and $J^{n}_{\text{zz}}{''}$). In addition it further distorts the octahedra and completely lifts the orbital degeneracy. We therefore expect  two of the three orbitals to be completely occupied giving rise to two strong antiferromagnetic exchange interactions, while the remaining orbital is unfilled resulting in no interaction. 
The issue is then to identify the unfilled orbital. In theory it should be possible to estimate the orbital energy level diagram by inspecting the octahedral environment as for the orthorhombic phase, however, this is difficult because the V-O distances are all unequal and the V ions are somewhat off-center \cite{Yan08,supp1}. Instead we turn to the magnetic structure to provide the answer. Figure \ref{fig:Pic1_zigzag_chains} shows that the magnetic moments coupled by $J^{n}_{\text{leg}}$ and $J^{n}_{\text{zz}}{'}$ have an antiparallel alignment while the moments coupled by $J^{n}_{\text{zz}}{''}$ are parallel.  This observation suggests that the leg exchange interaction ($J^{n}_{\text{leg}}$) and one of the zig-zag interactions ($J^{n}_{\text{zz}}{'}$) are strongly antiferromagnetic while the second zig-zag interaction ($J^{n}_{\text{zz}}{''}$) is weak. This in turn implies that the  $d^{n}_{\text{leg}}$ and $d^{n}_{\text{zz}}{'}$ orbitals are occupied, while the $d^{n}_{\text{zz}}{''}$ orbital is unoccupied as illustrated in Fig. \ref{fig:Pic4_mag_structure_bc}(b). Such an arrangement where every second interaction along the zig-zags is strong, combined with a strong leg exchange is a spin-1 ladder \cite{Mennerich06,Senechal95,Todo01}.\\\indent
To summarize, our investigation reveals the role played by orbital ordering in driving the spin-1, quasi-1D antiferromagnet CaV$_2$O$_4$ from one exotic spin Hamiltonian to another. High-temperature dc susceptibility combined with an inspection of the VO$_6$ octahedra reveal that in the orthorhombic phase there is partial orbital degeneracy and a single exchange interaction ($J_{\text{leg}}$) dominates. The system can therefore be viewed as a 1D, spin-1, Heisenberg antiferromagnet or Haldane chain. The orbital degeneracy is lifted in the monoclinic phase, long-range antiferromagnetic order develops and single-crystal neutron diffraction reveals a canted magnetic structure. The structure implies that the low temperature orbital configuration gives rise to a different exchange interaction and results in an antiferromagnetic, spin-1 ladder. There are to our knowledge only two other physical realizations of this system \cite{Mennerich06, Hosokoshi07}, both unlike CaV$_2$O$_4$, are organic. Theory suggests that the spin-1 ladder has gapped excitations but where the gap is considerably smaller than the Haldane gap ($\Delta_{\text{Haldane}} = 0.41J_{\text{leg}}$), e.g. for rung couplings in the range expected for CaV$_2$O$_4$ of $0.2J_{\text{leg}} \leq J_{\text{zz}}{'} \leq 1.0J_{\text{leg}}$, the gap is $0.16J_{\text{leg}} \geq \Delta_{\text{ladder(S=1)}} \geq 0.08J_{\text{leg}}$ \cite{Senechal95}. Assuming that the low-temperature value of $J_{\text{leg}}$ is similar to the high temperature values obtained from susceptibility (18 $\leq J_{\text{leg}} \leq$ 27 meV) - an assumption that is valid since the $d^{n}_{\text{leg}}$ orbital is occupied by one electron in both structural phases - the expected gap size should lie in the range 1.5 meV $\leq \Delta_{\text{ladder(S=1)}} \leq$ 4.4 meV. This small gap value explains why CaV$_2$O$_4$ is able to order below $T_{N}=$ 71 K in the presence of weak interladder coupling. Further work is in progress including band structure calculations to explore the low-temperature orbital splitting and inelastic neutron scattering to measure the magnetic excitation spectrum.\\\indent
We thank D. Khomskii and P.G. Radealli for their advice and R.J.\ McQueeney for supporting the crystal growth. M.R.\ and A.H.\ acknowledge funding from Deutsche Forschungsgemeinschaft (Grants No. UL 164/4 and No. HO 2325/4-1). Work at Ames was supported by the U.S. DOE (Contract No.~DE-AC02-07CH11358).
%\bibliographystyle{apsrev}
%\bibliography{cav2o4_new}

\end{document}